\documentclass[twocolumn,showpacs,preprintnumbers,amsmath,amssymb]{revtex4}

\usepackage{graphicx}
\usepackage{dcolumn}
\usepackage{bm}

\begin{document}


\title{Exact Quantum-Statistical Dynamics
of Time-Dependent Generalized Oscillators}

\author{Sang Pyo Kim}\email{sangkim@kunsan.ac.kr}

\affiliation{Department of Physics, Kunsan National University,
Kunsan 573-701, Korea}

\author{Don N. Page}\email{don@phys.ualberta.ca}
\affiliation{CIAR Cosmology Program, Theoretical Physics
Institute, Department of Physics, University of Alberta, Edmonton,
Alberta, Canada T6G 2J1}

\date{\today}
\begin{abstract}
Using linear invariant operators in a constructive way we find the
most general thermal density operator and Wigner function for
time-dependent generalized oscillators. The general Wigner
function has five free parameters and describes the thermal Wigner
function about a classical trajectory in phase space. The contour
of the Wigner function depicts an elliptical orbit with a constant
area moving about the classical trajectory, whose eccentricity
determines the squeezing of the initial vacuum.
\end{abstract}

\maketitle

A quantum system of time-dependent oscillators has been a
continuing issue of interest since the advent of quantum
mechanics. Paul trap is one of such oscillators, which has a
time-periodic frequency \cite{brown}. Recently geometric phase has
been studied for time-dependent quantum oscillators
\cite{jarzynski}. Various methods have been applied to
time-dependent quantum oscillators in many areas \cite{dekker}.
Agarwal and Kumar, and Aliaga {\it et al.} \cite{agarwal} studied
statistical properties of time-dependent oscillators. Also the
density matrix and density operator for time-dependent oscillators
were studied in Refs. \cite{eboli,kim-lee}.

On the other hand, Lewis and Riesenfeld \cite{lewis} introduced a
method to find the exact quantum states for the time-dependent
Schr\"{o}dinger equation. In particular, for time-dependent
oscillators they found a quadratic invariant operator, satisfying
the quantum Liouville-von Neumann equation, whose eigenstates
provide the exact quantum states up to time-dependent phase
factors. Even for time-dependent generalized oscillators each
complex solution to the classical equation of motion leads to a
pair of invariant operators, linear in position and momentum
operators \cite{kim-page}.

In this paper, using the linear invariant operators, we find in a
constructive way the most general thermal density operator and
Wigner function up to the quadratic order in position and momentum
operators. This density operator is a squeezed and displaced state
of a thermal one. Further, the density matrix is the thermal one
shifted by a real classical solution, which has five free
parameters. The contour of the Wigner function follows an
elliptical orbit with a constant area whose center moves and
principal axes rotate along a classical trajectory. The shape of
the ellipse measured by eccentricity determines the squeezing of
the initial vacuum.

The time-dependent generalized quantum oscillator is described by
the Hamiltonian \cite{cervero,kim-page}
\begin{equation}
\hat{H} (t) = \frac{X(t)}{2} \hat{p}^2 + \frac{Y(t)}{2}
(\hat{p}\hat{q} + \hat{q}\hat{p}) + \frac{Z(t)}{2} \hat{q}^2,
\label{osc}
\end{equation}
where $X, Y$ and $Z$ explicitly depend on time. Lewis and
Riesenfeld have shown that the invariant operator satisfying the
quantum Liouville-von Neumann equation
\begin{equation}
i \hbar \frac{\partial}{\partial t} \hat{I} (t) + [\hat{I} (t),
\hat{H} (t)] = 0, \label{ln eq}
\end{equation}
provides the exact quantum states of the time-dependent
Schr\"{o}dinger equation as its eigenstates up to time-dependent
phase factors.  Following Ref. \cite{kim-page} we introduce a pair
of linear invariant operators
\begin{eqnarray}
\hat{a}_u (t)  &=& \frac{i}{\sqrt{\hbar}} \Biggl[u^*(t) \hat{p} -
\frac{1}{X(t)} [ \dot{u}^* (t) - Y(t) u^* (t)] \hat{q} \Biggr],
\nonumber\\ \hat{a}^{\dagger}_u (t) &=& - \frac{i}{\sqrt{\hbar}}
\Biggl[u (t) \hat{p} - \frac{1}{X(t)} [ \dot{u} (t) - Y(t) u (t)]
\hat{q} \Biggr], \label{inv op}
\end{eqnarray}
where $u$ is a complex solution to the classical equation of
motion
\begin{equation}
\frac{d}{dt} \Biggl(\frac{\dot{u}}{X} \Biggr) + \Biggl[ XZ - Y^2 +
\frac{\dot{X}Y - X \dot{Y}}{X}\Biggr] \Biggl(\frac{u}{X} \Biggr) =
0. \label{cl eq}
\end{equation}
with overdots denoting the derivative with respect to $t$.
Normalizing the complex solution to satisfy the Wronskian
condition
\begin{equation}
{\rm Wr} \{u^*, u\} = \frac{1}{X} (u \dot{u}^* - u^* \dot{u}) = i,
\label{wr}
\end{equation}
one can make the invariant operators satisfy the standard
commutation relation
\begin{equation*}
[\hat{a}_u (t), \hat{a}^{\dagger}_u (t) ] = 1.
\end{equation*}

Another complex solution $v$ to Eq. (\ref{cl eq}), which can be
expressed as a linear superposition of $u$:
\begin{eqnarray*}
v(t) &=& \mu^* u (t) - \nu^* u^* (t),
\end{eqnarray*}
for complex constants $\mu$ and $\nu$ given by
\begin{equation*}
\mu = i {\rm Wr} \{u, v^* \}, \quad \nu =  i{\rm Wr} \{u^*, v^*\},
\end{equation*}
leads to another set of the invariant operators $\hat{a}_v$ and
$\hat{a}^{\dagger}_v$. The Wronskian condition on $v$
\begin{equation*}
{\rm Wr}\{v^*, v\} = i \Leftrightarrow |\mu|^2 - |\nu|^2 = 1,
\end{equation*}
also guarantees the standard commutation relation
\begin{equation*}
[\hat{a}_v (t), \hat{a}^{\dagger}_v (t)] = 1.
\end{equation*}
In fact, these two sets of invariant operators are related through
the Bogoliubov transformation
\begin{eqnarray*}
\hat{a}_v (t) &=& \mu \hat{a}_u (t) + \nu \hat{a}^{\dagger}_u (t),
\nonumber\\ \hat{a}^{\dagger}_v (t) &=& \mu^* \hat{a}^{\dagger}_u
(t) + \nu^* \hat{a}_u (t). \label{bog}
\end{eqnarray*}
The Bogoliubov transformation is written as the similarity
transform \cite{yuen,agarwal}
\begin{equation*}
\hat{a}_v (t) = \hat{S}^{-1} (t) \hat{a}_u (t) \hat{S} (t), \quad
\hat{a}^{\dagger}_v (t) = \hat{S}^{-1} (t) \hat{a}^{\dagger}_u (t)
\hat{S} (t),
\end{equation*}
by the squeezing operator
\begin{equation*}
\hat{S} (t) = e^{ i \theta_{\mu} \hat{a}^{\dagger}_u \hat{a}_u}
\exp \Biggl[\frac{1}{2}e^{i(\theta_{\nu} - \theta_{\mu})}
\cosh^{-1}|\mu| \hat{a}^{\dagger 2}_u - {\rm H.c} \Biggr],
\end{equation*}
where $\mu = |\mu| e^{i \theta_{\mu}}, \nu = |\nu| e^{i
\theta_{\nu}}$. We may use the freedom in choosing the overall
constant phase of $\hat{a}_v$, which is not physically important,
to fix the phase $\theta_{\mu} = 0$ \cite{yuen}. Thus there are
only two parameters $|\mu|$ and $\theta_{\nu}$ or a complex
constant $\nu$, i.e. $|\nu|$ and $\theta_{\nu}$.

The most general, quadratic, Hermitian invariant operator
constructed from the pair $\hat{a}_v$ and $\hat{a}^{\dagger}_v$
takes the form
\begin{eqnarray}
\hat{\cal I}_v (t) &=& \frac{A}{2} \hat{a}^{\dagger2}_v (t) +
\frac{B}{2} [\hat{a}^{\dagger}_v (t) \hat{a}_v (t) + \hat{a}_v (t)
\hat{a}^{\dagger}_v (t)] + \frac{A^*}{2} \hat{a}^2_v (t)
\nonumber\\
&& +  D \hat{a}^{\dagger}_v (t) + D^* \hat{a}_v (t) + E,
\label{gen quad}
\end{eqnarray}
where $A$ and $D$ are complex constants, and $B$ and $E$ are real
constants. By choosing $\mu$ and $\nu$, i.e. $u$ such that
\begin{equation*}
A \mu^{*2} + 2 B \mu^* \nu + A^* \nu^{2} = 0,
\end{equation*}
the invariant operator (\ref{gen quad}) can be written in the
canonical form
\begin{eqnarray}
\hat{\cal I}_u (t) = \hbar \omega_0 \hat{a}^{\dagger}_u (t)
\hat{a}_u (t) + \delta \hat{a}^{\dagger}_{u} (t) + \delta^*
\hat{a}_u (t) + \epsilon. \label{can form}
\end{eqnarray}
where
\begin{eqnarray*}
\hbar \omega_0 &=& A \mu^* \nu^* + B (|\mu|^2 + |\nu|^2) + A^* \mu
\nu, \nonumber\\
\delta &=& D \mu^* + D^* \nu, \nonumber\\
\epsilon &=& E + \frac{1}{2}\hbar \omega_0.
\end{eqnarray*}
Hence this implies that by allowing all the complex $u$'s
satisfying both Eqs. (\ref{cl eq}) and (\ref{wr}) the invariant
operator (\ref{can form}) is general enough for our purpose. From
now on we shall work on the Fock bases $\hat{a}_u$ and
$\hat{a}^{\dagger}_u$ for all the complex $u$'s and drop the
subscript $u$.

Since the invariant operator (\ref{can form}) satisfies Eq.
(\ref{ln eq}), we use it to define the density operator
\cite{kim-lee}
\begin{equation}
\hat{\rho} (t) = \frac{1}{Z} e^{- \beta \hat{\cal I}_u (t)}.
\label{den op}
\end{equation}
Here $\beta$ is a free parameter that may be identified with the
inverse temperature of the system in equilibrium, and $Z = {\rm
Tr} (e^{- \beta \hat{\cal I}_u})$. The density operator has five
free parameters, i.e. $\beta$ or $\omega_0$, a complex constant
$\delta$, which is related with the classical position $q_c$ and
momentum $p_c$ as will be shown below, and $|\mu|$ and
$\theta_{\nu}$ in choosing $u$. By introducing the displacement
operator
\begin{equation*}
\hat{D} (z) = e^{- z \hat{a}^{\dagger}(t) + z^* \hat{a} (t)},
\end{equation*}
with $z = - \delta/(\hbar \omega_0)$, $\epsilon =
|\delta|^2/(\hbar \omega_0)$, we write the density operator as
\begin{equation}
\hat{\rho} (t) = \hat{D}^{\dagger} (z) \hat{\rho}_{\rm T} (t)
\hat{D} (z), \label{dis den}
\end{equation}
where
\begin{equation*}
\hat{\rho}_{\rm T} (t) = \frac{1}{Z_{\rm T}} e^{- \beta \hbar
\omega_0 \hat{a}^{\dagger} (t) \hat{a} (t)},
\end{equation*}
is a thermal density operator. It follows that $Z = Z_{\rm T}$ due
to the unitary transformation (\ref{dis den}). The coherent state,
defined as $\hat{a} (t) \vert z,t \rangle = z \vert z, t \rangle$,
is also given by
\begin{equation*}
\vert z, t \rangle = \hat{D}^{\dagger} (z) \vert 0, t \rangle,
\end{equation*}
where $\vert 0, t \rangle $ is the vacuum state that is
annihilated by $\hat{a} (t)$. The position and momentum
expectation value with respect to the coherent state is
\begin{eqnarray}
\langle z, t \vert \hat{q} \vert z, t \rangle &=& \sqrt{\hbar} (u
z + u^* z^*) \equiv q_c, \nonumber\\
\langle z, t \vert \hat{p} \vert z, t \rangle &=& - \frac{Y}{X}
q_c + \frac{\sqrt{\hbar}}{X} (\dot{u} z + \dot{u}^* z^*) \equiv
p_c. \label{cl pm}
\end{eqnarray}
The $q_c$ and $p_c$ satisfy the classical Hamilton equations
\begin{eqnarray*}
\dot{q}_c &=& X p_c + Y q_c, \nonumber\\
\dot{p}_c &=& - Y p_c - Z q_c.
\end{eqnarray*}

Now, from the definition of the thermal expectation value
\begin{equation*}
\langle \hat{\cal O} \rangle = {\rm Tr} [\hat{\cal O}
\hat{\rho}(t)] = {\rm Tr} [ \hat{D}(z) \hat{\cal O}
\hat{D}^{\dagger} (z) \hat{\rho}_{\rm T} ],
\end{equation*}
we find the expectation value of position and momentum operators
\begin{eqnarray*}
\langle \hat{q} \rangle &=&  q_c, \quad \langle \hat{p} \rangle =
p_c,
\end{eqnarray*}
and that of quadratic operators
\begin{eqnarray*}
\langle \hat{q}^2  \rangle &=& \hbar u^* u (1 + 2
\bar{n}) + q_c^2, \nonumber\\
\langle \hat{p}^2  \rangle &=& \frac{\hbar}{X^2}
(\dot{u}^* - Y u^*) (\dot{u} - Y u) (1 + 2 \bar{n}) + p_c^2, \nonumber\\
\langle \frac{1}{2} (\hat{q} \hat{p} + \hat{p} \hat{q}) \rangle
&=& \frac{\hbar}{2X} [(\dot{u}^* - Y u^*)u + (\dot{u} - Y u) u^* ]
\nonumber\\ && \times (1 + 2 \bar{n})  + q_c p_c.
\end{eqnarray*}
where
\begin{equation*}
\bar{n} = \frac{1}{ e^{\beta \hbar \omega_0} - 1}
\end{equation*}
is the mean number density of Bose-Einstein distribution. The
vacuum result is obtained by taking the limit $\beta \rightarrow
\infty$. It is worth noting that the dispersion of position and
momentum around the classical trajectory $(q_c, p_c)$ is entirely
determined by the thermal one: $\langle (\hat{q} - q_c)^2 \rangle
= \langle \hat{q}^2 \rangle_{\rm T}$ and $\langle (\hat{p}- p_c)^2
\rangle = \langle \hat{p}^2 \rangle_{\rm T}$, where $\langle
\hat{\cal O} \rangle_{\rm T} = {\rm Tr} [\hat{\cal O}
\hat{\rho}_{\rm T}(t)]$.

Using $ \hat{D} (z) = e^{i p_c q_c/2 \hbar} e^{i q_c
\hat{p}/\hbar} e^{- i p_c \hat{q}/\hbar}$, we find the coordinate
representation of the displacement operator
\begin{eqnarray*}
\langle q \vert \hat{D} (z) \vert q' \rangle =  e^{i p_c q_c/2
\hbar} e^{- i p_c q'/\hbar} \delta (q + q_c - q'), \label{d rep}
\end{eqnarray*}
and that of its Hermitian conjugate, $ \langle q \vert
\hat{D}^{\dagger} (z) \vert q' \rangle = \langle q' \vert \hat{D}
(z) \vert q \rangle^*$. Hence the density matrix is given by
\begin{eqnarray}
\rho (q, q') &=& \langle q \vert \hat{\rho} (t) \vert q' \rangle
\nonumber\\
 &=& \langle q \vert \hat{D}^{\dagger}
(z) \int dq_1 \vert q_1 \rangle \langle q_1 \vert \hat{\rho}_{\rm
T} \int dq_2 \vert q_2 \rangle \langle q_2 \vert \hat{D} (z) \vert
q' \rangle
\nonumber\\
&=& e^{i p_c (q - q')/\hbar} \rho_{\rm T} (q - q_c, q' - q_c, t),
\label{den mat}
\end{eqnarray}
where $\rho_{\rm T}$ is the density matrix for the thermal state
given, for instance, in Ref. \cite{kim-lee}. Finally, the Wigner
function is given by
\begin{eqnarray}
P (q, p) &=& \frac{1}{\pi \hbar} \int_{- \infty}^{\infty} dy
\rho (q-y, q+y) e^{2i py/\hbar} \nonumber\\
&=& P_{\rm T} (q - q_c, p - p_c), \label{wig}
\end{eqnarray}
where $P_{\rm T}$ is the Wigner function for the thermal state:
\begin{eqnarray*}
P_{\rm T} (q, p) &=& \frac{\tanh(\beta \hbar \omega_0/2)}{\pi
\hbar} \exp \Biggl[- \frac{2 \tanh(\beta \hbar \omega_0/2) }{\hbar
\omega_0 } {\cal H}_{\rm E} \Biggr], \nonumber\\
{\cal H}_{\rm E} (q, p) &=& \omega_0 u^* u \Biggl(p - \frac{d \ln
(u^*u)^{1/2}}{dt} \frac{q}{X} \Biggr)^2 + \frac{\omega_0}{4 u^*u}
q^2.
\end{eqnarray*}
The Wigner function (\ref{wig}) also has five parameters: $q_c$,
$p_c$, $\beta$ (or $\omega_0$), $\mu$, and $\theta_{\nu}$.

The Wigner functions $P$ and $P_{\rm T}$ and their vacuum limit
are positive definite in contrast with those for excited states
that may take negative values in some region of phase space
\cite{kim-lee2}. Hence $P$ and $P_{\rm T}$ may be used as a
distribution of phase space for the quantum evolution. For
instance, the harmonic oscillator with $X = 1/X, Y = 0$ and $Z =
m_0 \omega_0^2$, has the solution $u (t) = e^{- i \omega_0
t}/\sqrt{2 m_0 \omega_0}$ which recovers the well-known Wigner
function with ${\cal H}_{\rm E} (q, p) = H (q, p)$. In general,
${\cal H}_{\rm E}$ depicts an ellipse centered at the origin,
which can be written in the canonical form
\begin{eqnarray*}
{\cal H}_{\rm E} &=& \Bigl(\lambda_+ \tilde{p}^2 + \lambda_-
\tilde{q}^2 \Bigr) \times \frac{\omega_0}{2},
\nonumber\\
\lambda_{\pm} &=& u^* u + \frac{1}{4 u^* u} +
\frac{u^* u }{X} \Bigl(\frac{d \ln (u^*u)^{1/2}}{dt} \Bigr)^2 \\
&& \pm \Biggl[\Biggl\{ u^* u + \frac{1}{4 u^* u} + \frac{u^* u
}{X} \Bigl(\frac{d \ln (u^*u)^{1/2}}{dt} \Bigr)^2 \Biggr\}^2 - 1
\Biggr]^{1/2},
\end{eqnarray*}
where $(\tilde{q}, \tilde{p})$ are new phase-space coordinates
rotated with the angle
\begin{equation*}
\tan [2 \theta (t)] = \frac{2(u^* u/X) [d \ln
(u^*u)^{1/2}/dt]}{u^* u  + (1/4 u^* u) - (u^*u/X^2) [d \ln
(u^*u)^{1/2}/dt]^2}.
\end{equation*}
Since $\lambda_+ \lambda_- = 1$, the area of the ellipse does not
depend on the solution $u$. Therefore, as shown in Fig. 1, the
contour of the Wigner function (\ref{wig}) follows an elliptical
orbit with a constant area whose center $(q_c, p_c)$ in turn moves
on a classical trajectory and principal axes $(\tilde{q},
\tilde{p})$ rotate with the angle $\theta (t)$. The shape of the
ellipse somehow determines the squeezing of the initial vacuum
\cite{kim-noz}.

Now we introduce a geometric measure for the squeezing of the
initial vacuum and particle production in terms of the
eccentricity of the ellipse. For that purpose we consider the
exactly solvable oscillator \cite{birrel,kim-lee}
\begin{equation*}
X = \frac{1}{m}, \quad Y = 0, \quad Z = m [\omega_1^2 - \omega_0^2
\tanh(t/ \tau)].
\end{equation*}
The oscillator has an asymptotic frequency $\omega_i = (\omega_1^2
+ \omega_0^2)^{1/2}$ at $t = - \infty$ and $\omega_f = (\omega_1^2
- \omega_0^2)^{1/2}$ at $t = \infty$. As $\tau$ is an interval for
the frequency change, the adiabatic change is prescribed by the
condition $\tau \gg 1$. The solution that has the correct
asymptotic form $u = e^{- i \omega_i t} / \sqrt{2 m \omega_i}$ at
$t = - \infty$ is given by
\begin{equation*}
u (t) = \frac{e^{- i \omega_i t}}{\sqrt{2 m \omega_i}} {}_2 F_1 (-
i \frac{\tau}{2} (\omega_i + \omega_f), - i \frac{\tau}{2}
(\omega_i - \omega_f); 1 - i \tau \omega_i; - e^{2 t/\tau}),
\end{equation*}
where ${}_2 F_1$ is the hypergeometric function. At $t = - \infty$
we obtain ${\cal H}_{\rm E} = p^2/(2m) + m \omega_i^2 q^2/2 =
H_i$. Whereas, at $t = \infty$ the solution has another asymptotic
form
\begin{equation*}
u (t) = \frac{1}{\sqrt{2 m \omega_i}} [ \alpha_{+} e^{ - i
\omega_f t} + \alpha_{-} e^{ i \omega_f t}],
\end{equation*}
where
\begin{equation*}
\alpha_{\pm} (\tau) = \frac{\Gamma (1 - i \omega_i \tau)
\Gamma(\mp i \omega_f \tau)}{\Gamma(1 - i \frac{\tau}{2}(\omega_i
\pm \omega_f)) \Gamma(- i \frac{\tau}{2} (\omega_i \pm
\omega_f))}.
\end{equation*}
For the adiabatic limit $\tau \rightarrow  \infty$, $|\alpha_-|
\rightarrow 0$ and  $u (t = \infty) = \alpha_+ e^{ - i \omega_f
t}/\sqrt{2 m \omega_i}$. As $|\alpha_+ (\tau = \infty)| =
\sqrt{\omega_i/ \omega_f}$, we get ${\cal H}_{\rm E} = p^2/(2m) +
m \omega_f^2 q^2/2 = H_f$. The contour of the Wigner function at
$t = - \infty$ is given by an ellipse centered at the classical
position and momentum
\begin{equation*}
q_c = q_i \cos (\omega_i t + \varphi_i), \quad p_c = - m \omega_i
q_i \sin (\omega_i t + \varphi_i),
\end{equation*}
where
\begin{eqnarray*}
q_i = \sqrt{\frac{2 \hbar}{m \omega_i}} |z|, \quad e^{- i
\varphi_i} = \frac{z}{|z|},
\end{eqnarray*}
and at $t = \infty$, centered at
\begin{equation*}
q_c = q_f \cos (\omega_f t + \varphi_f), \quad p_c = - m \omega_f
q_f \sin (\omega_f t + \varphi_f),
\end{equation*}
where
\begin{eqnarray*}
q_f = \sqrt{\frac{2 \hbar}{m \omega_i}} | \alpha_+ z|, \quad e^{-
i \varphi_f} = \frac{\alpha_+ z}{|\alpha_+  z|}.
\end{eqnarray*}
From the Bogoliubov transformation between the invariant operators
(\ref{inv op}) for $H_i$ and $H_f$ follows the complex squeezing
parameter
\begin{eqnarray*}
\nu &=& i m [ \dot{u}^*_i (t) u^*_f (t) - \dot{u}^*_f (t) u^*_i (t)]\\
&=& \frac{1}{2} \Biggl(\sqrt{\frac{\omega_f}{\omega_i}} -
\sqrt{\frac{\omega_i}{\omega_f}} \Biggr)e^{i (\omega_i +
\omega_f)t}.
\end{eqnarray*}
In terms of the eccentricity $e_i$ and $e_f$ of $H_i$ and $H_f$ we
obtain the expression
\begin{equation*}
| \nu | = \frac{|(1 - e_i^2)^{1/2} - (1 - e_f^2)^{1/2}|}{2[(1 -
e_i^2)(1 - e_f^2)]^{1/4}}.
\end{equation*}
Thus we show that the geometric shape determined by the
eccentricity of the ellipse measures the squeezing of the vacuum
state and thereby the amount of particle production.

As an illustrative but nontrivial application of our general
Wigner function, we consider the harmonic oscillator with
\begin{equation*}
X = \frac{1}{m}, \quad Y = 0, \quad Z = m \omega_0^2,
\end{equation*}
which has the most general solution
\begin{equation*}
u (t) = \frac{1}{\sqrt{2 m \omega_0}} [\mu e^{- i \omega_0 t} +
\nu e^{i \omega_0 t}], \quad (\nu \neq 0).
\end{equation*}
As shown in Fig. 1, the contour of the Wigner function moves on a
small elliptical orbit about another elliptical orbit for the
classical trajectory
\begin{equation*}
q_c = q_0 \cos (\omega_0 t + \varphi_0), \quad p_c = - m \omega_0
q_0 \sin (\omega_0 t + \varphi_0),
\end{equation*}
where
\begin{eqnarray*}
q_0 = \sqrt{\frac{2 \hbar}{m \omega_0}} |\mu z + \nu^* z^*|, \quad
e^{- i \varphi_0} = \frac{\mu z + \nu^* z^*}{|\mu z + \nu^* z^*|}.
\end{eqnarray*}
In fact, the contour is similar to the epicycle of an elliptical
orbit moving on another elliptical orbit. Now the eccentricity
$e(t) = \sqrt{{\rm min} \{\lambda_{\pm} (t) \}/ {\rm max}
\{\lambda_{\pm} (t) \}}$ and the rotation angle $\theta (t)$ of
the small elliptical orbit explicitly depend on time through the
magnitude of the complex solution
\begin{equation*}
u^* u  = |\mu|^2 + |\nu|^2 + \mu \nu^* e^{- 2 i \omega_0 t} +
\mu^* \nu e^{2 i \omega_0 t}.
\end{equation*}
Hence the density matrix and Wigner function for $\nu \neq 0$
describe various kinds of nontrivial quantum states beyond the
standard static ones.

In summary, we showed that the most general, quadratic, Hermitian
invariant operator (\ref{gen quad}) can be written in the
canonical form (\ref{can form}) by suitably choosing a complex
solution $u$ to the classical equation of motion (\ref{cl eq}).
Then the general thermal density operator takes the form (\ref{den
op}), which is nothing but the displaced state of the thermal
state (\ref{dis den}). The density matrix (\ref{den mat}) and
Wigner function (\ref{wig}), the main results of this paper, are
the thermal ones shifted by a classical solution. The general
Wigner function for thermal state has five free parameters, which
characterize the classical configuration $(q_c, p_c)$ together
with $\beta$ (or $\omega_0$) for thermal equilibrium, $|\mu|$ (or
$|\nu|$) for the squeezing of the initial vacuum, and
$\theta_{\nu}$. Also it is positive definite and describes a
distribution of phase space for quantum evolution. Further, we
showed that the contour of the Wigner function depicts an
elliptical orbit centered at a classical trajectory. Our general
Wigner function provides a nontrivial time-dependent Wigner
function even for a time-independent harmonic oscillator, whose
contour describes an analog of epicycle of an elliptical orbit
moving about another elliptical orbit.

\acknowledgements

The work of S.P.K. was supported by the Korea Research Foundation
under Grant No. 2000-015-DP0080 and the work of D.N.P. by the
Natural Sciences and Engineering Council of Canada.

\begin{figure}
\includegraphics{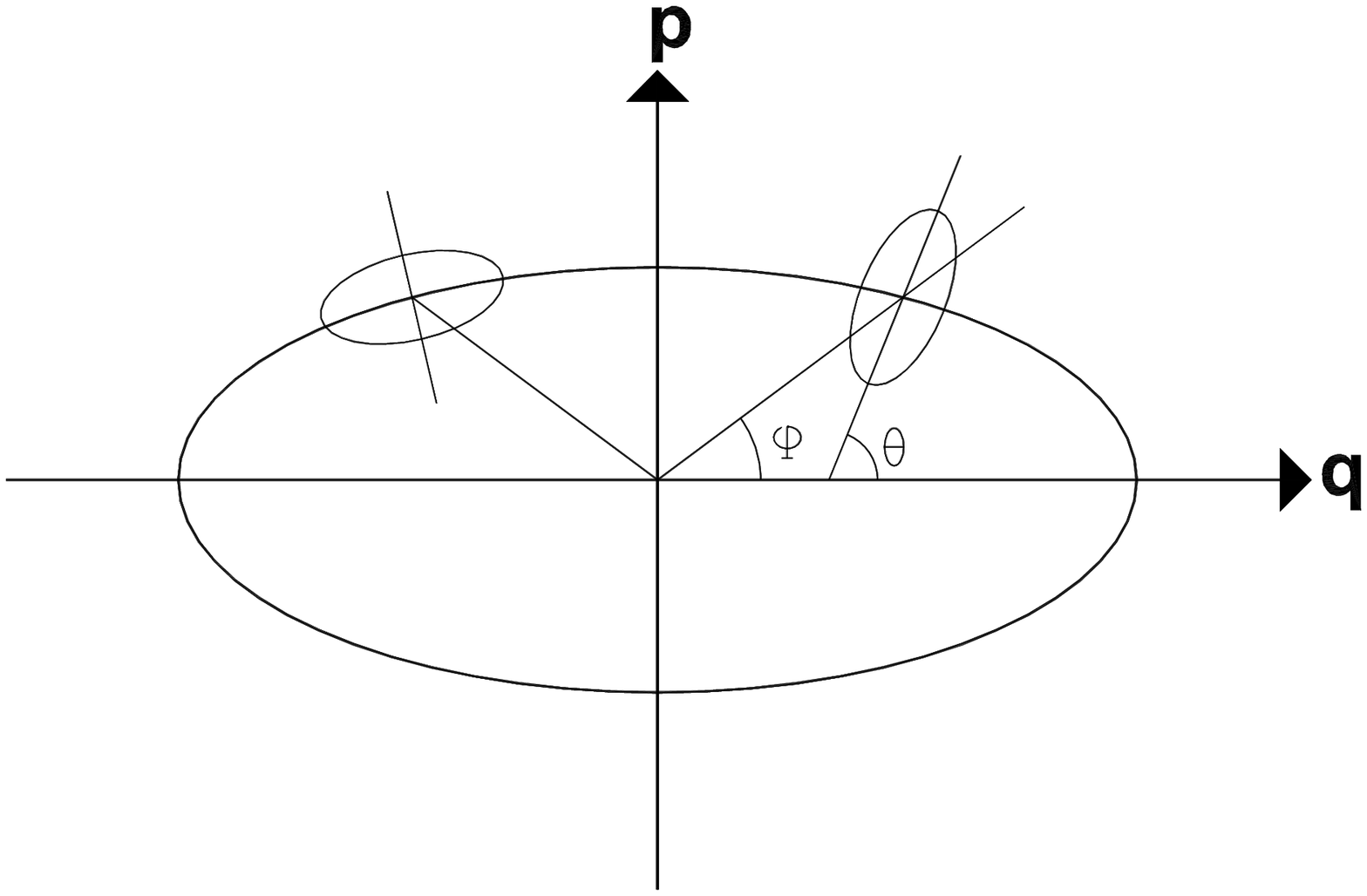}
\caption{The contour of the Wigner function $P$. The ellipse
centered at the origin describes the classical trajectory, a
periodic motion, and those centered on the ellipse correspond to a
thermal state. The $\varphi$ is the polar angle of the classical
position $(q_c, p_c)$.}
\end{figure}

\end{document}